\RequirePackage{fix-cm}
\documentclass[smallextended]{svjour3}       
\smartqed  
\usepackage{graphicx}
\usepackage{amssymb}
\usepackage[ruled,vlined,linesnumbered]{algorithm2e}
\SetEndCharOfAlgoLine{}

\usepackage{color}

\usepackage{amsmath,amsfonts,amssymb}
\usepackage{mathtools}
\usepackage{multicol}

\newcommand{\FF}{\ensuremath{\mathbb{F}}}
\newcommand{\FFbar}{\ensuremath{\overline{\FF}}}
\newcommand{\OO}{\ensuremath{O}}
\newcommand{\PP}{\ensuremath{\mathbb{P}}}
\newcommand{\ZZ}{\ensuremath{\mathbb{Z}}}

\newcommand{\EC}{\ensuremath{\mathcal{E}}}
\newcommand{\HC}{\ensuremath{\mathcal{H}}}

\newcommand{\subgrp}[1]{\ensuremath{\langle{#1}\rangle}}
\newcommand{\xMap}{\ensuremath{\mathbf{x}}}

\newcommand{\FqM}{\ensuremath{\mathtt{M}}}
\newcommand{\FqMconst}{\ensuremath{\mathtt{c}}}
\newcommand{\FqS}{\ensuremath{\mathtt{S}}}
\newcommand{\Fqa}{\ensuremath{\mathtt{a}}}
\newcommand{\Fqs}{\ensuremath{\mathtt{s}}}
\newcommand{\xDBL}{{\tt xDBL}}
\newcommand{\xADD}{{\tt xADD}}
\newcommand{\LADDER}{{\tt LADDER}}
\newcommand{\EUCLID}{{\tt EUCLID2D}}
\newcommand{\PRAC}{{\tt PRAC}}
\newcommand{\CSWAP}{\ensuremath{\mathtt{SWAP}}} 
\newcommand{\Recover}{{\tt Recover}}

\begin{document}

\title{Montgomery curves and their arithmetic
}
\subtitle{The case of large characteristic fields}


\author{Craig Costello        \and
        Benjamin Smith 
}

\dedication{A survey in tribute to Peter L. Montgomery}


\institute{Craig Costello \at
              Microsoft Research, USA \\
              \email{craigco@microsoft.com}           
           \and
           Benjamin Smith \at
             INRIA \emph{and} Laboratoire d'Informatique de l'\'{E}cole polytechnique (LIX), 
    France \\
    \email{smith@lix.polytechnique.fr}  
}

\maketitle

\begin{abstract}
    Three decades ago, Montgomery introduced a new elliptic curve
    model for use in Lenstra's ECM factorization algorithm.  
    Since then, his curves
    and the algorithms associated with them have become foundational
    in the implementation of elliptic curve cryptosystems.
    This article surveys the theory and cryptographic applications of Montgomery curves 
    over non-binary finite fields,
    including Montgomery's \(x\)-only arithmetic
    and Ladder algorithm, \(x\)-only Diffie--Hellman,
    \(y\)-coordinate recovery, and 2-dimensional and Euclidean
    differential addition chains such as Montgomery's PRAC algorithm.
\end{abstract}

\section{
    Introduction
}

Peter L.~Montgomery's landmark 1987 paper
\emph{Speeding the Pollard and elliptic curve methods of
factorization}~\cite{montgomery87}
introduced what became known as \emph{Montgomery curves}
and the \emph{Montgomery ladder}
as a way of accelerating Lenstra's ECM factorization method~\cite{lenstra87}.
However,
they have gone on to have a far broader impact:
while remaining a crucial component of modern factoring software,
they have also become central to elliptic curve cryptography.

Consider the following situation:
let \(q\) be a prime power and let \(\EC\) be an elliptic curve over
\(\FF_q\),
with group law \(\oplus\), identity point \(\OO\),
and negation map~\(\ominus\). 
We have \emph{scalar multiplication} endomorphisms on \(\EC\)
defined by
\[
    [k] : P 
    \longmapsto
    \underbrace{P\oplus\cdots\oplus P}_{k \text{ times}}
\]
for every \(k\) in \(\ZZ\) 
(with \([-k]P := [k](\ominus P)\)).
The negation map \(\ominus\) is an automorphism of \(\EC\),
and the quotient of \(\EC\) by \(\subgrp{\ominus}\)
is the 2-to-1 mapping
\[
    \xMap: \EC \longrightarrow \PP^1 \cong \EC/\subgrp{\ominus} 
\]
such that 
\(\xMap(P) = \xMap(Q)\) if and only if \(P = Q\) or \(P=\ominus Q\).
If \(\EC\) is defined by a Weierstrass equation 
\(y^2 = f(x)\), then \(\xMap\) is just \(P \mapsto x(P)\);
in other models or coordinate systems, \(\xMap\) may be more complicated.
We call \(\PP^1\) the \emph{\(x\)-line} of \(\EC\).

Now, \(\PP^1\) does not inherit any group structure from \(\EC\).
But since \(\ominus\) commutes with \([k]\)
(i.e., \([k]({\ominus P}) = {\ominus[k]P}\)),
we still get an induced \emph{pseudomultiplication}
\[
    \xMap(P) \longmapsto \xMap([k]P)
\]
on \(\PP^1\) for every \(k\) in \(\ZZ\).

\label{sec:x-only-DH-intro}
In his seminal article proposing elliptic curves for use in cryptography, 
Miller~\cite{miller85} pointed out that elliptic curve Diffie--Hellman 
key exchange can be expressed entirely in terms of the maps
\(\xMap(P) \mapsto \xMap([k]P)\):
given a public base point \(\xMap(P)\) on \(\PP^1\), 
Alice (resp.~Bob) can compute and publish $\xMap([a]P)$ 
(resp.~\(\xMap([b]P)\))
for some secret \(a\) and \(b\),
and then derive the shared secret \(\xMap([b][a]P) = \xMap([a][b]P)\)
from \(\xMap([b]P)\) (resp.~\(\xMap([a]P)\)).
We can always lift the resulting Diffie--Hellman problem to \(\EC\),
so \(\EC\) provides not only the map \(\xMap\),
but also the security of the whole protocol;
but by working in \(\PP^1(\FF_q)\) instead of \(\EC(\FF_q)\)
we can save some space, and (hopefully) some time.

As a second application, consider ECM.
We are given an integer \(N\) whose prime factorization is unknown.
Choosing a random elliptic curve \(\EC\) over \(\ZZ/N\ZZ\)
and constructing a point \(P\) in \(\EC(\ZZ/N\ZZ)\),
we can compute \([B!]P\) for some moderate bound \(B\);
if there exists a prime factor \(p\) of \(N\)
such that \(\#\EC(\FF_p)\) is \(B\)-smooth,
then we will have \([B!]P \equiv \OO \pmod{p}\);
and we can detect this situation by taking the GCD of the projective
\(Z\)-coordinate of \([B!]P\) (for a Weierstrass model of \(\EC\))
with \(N\). 
If the GCD is neither \(1\) nor \(N\),
then we have found a factor of \(N\);
otherwise, we can try again with another random \(\EC\).
But all this still holds if we replace \(P\) with \(\xMap(P)\)
and \([B!]P\) with \(\xMap([B!]P)\);
and the advantages of compactness and simplicity that we can see in
Diffie--Hellman on \(\PP^1\) are magnified in the ECM context,
where the integer \(N\) can be much larger 
than any reasonable cryptographic \(q\).

In both scenarios, our task is simple:
we need to define a class of curves \(\EC\)
equipped with efficient algorithms for computing 
\(\xMap(P) \mapsto \xMap([k]P)\) on \(\PP^1\).

Montgomery's work provides a brilliant answer to this problem.
The \emph{Montgomery ladder} is a simple yet efficient
algorithm for computing \(\xMap(P) \mapsto \xMap([k]P)\).
In the abstract, the ladder is a sequence of steps
of pseudo-operations derived from the curve \(\EC\)
(see~\S\ref{sec:pseudo-ops});
choosing a \emph{Montgomery curve} for \(\EC\) 
ensures that each of those steps is optimized.
The result, for Montgomery,
was an extremely efficient implementation of ECM;
and even today, three decades later,
Montgomery's methods remain at the heart of 
state-of-the-art factoring software
such as the widely-distributed \texttt{GMP-ECM} package~\cite{GMP-ECM,ECM20}.
Later, these same qualities also led to many efficient
implementations of elliptic curve cryptosystems,
most notably Bernstein's Curve25519 software~\cite{curve25519}.

\paragraph{Notation.}
Throughout,
we work over the finite field \(\FF_q\),
where \(q\) is a power of an odd prime \(p\)
(for most contemporary applications, \(q = p\) or \(p^2\)).
We write \(\FqM\), \(\FqS\), \(\Fqa\), and~\(\Fqs\) 
for the cost of a single multiplication, squaring, addition, and
subtraction in \(\FF_q\), respectively.
We will occasionally need to multiply by some constant elements of
\(\FF_q\), which we hope to make as cheap as possible:
we write \(\FqMconst\) for the cost of a single such multiplication,
to help keep track of this cost separately from the other
multiplications with two variable inputs.

\section{
    Montgomery curves
}
\label{sec:curves}


A \emph{Montgomery curve} over \(\FF_q\) 
is an elliptic curve defined by an affine
equation
\[
    \EC_{(A,B)}: By^2 = x(x^2 + Ax + 1) 
    \ ,
\]
where \(A\) and \(B\) are parameters in \(\FF_q\)
satisfying\footnote{%
    These conditions imply nonsingularity:
    if \(B = 0\) then \(\EC_{(A,B)}\) is a union of three lines,
    while if \(A^2 = 4\) then \(\EC_{(A,B)}\) is a nodal cubic.
}
\(B \not= 0\) and \(A^2 \not= 4\).
Moving to projective plane coordinates \((X:Y:Z)\),
with 
\[
    x = X/Z
    \quad
    \text{and}
    \quad
    y = Y/Z
    \ ,
\]
we have the projective model
\begin{equation}
    \label{eq:defining-equation}
    \EC_{(A,B)}: BY^2Z = X(X^2 + AXZ + Z^2)
    \subseteq 
    \PP^2
    \ .
\end{equation}
There is a unique point \(\OO = (0:1:0)\) at infinity on \(\EC_{(A,B)}\):
it is the only point on \(\EC_{(A,B)}\) where \(Z = 0\).

\subsection{The parameters \(A\) and \(B\)}\label{sub:paramsAB}

A Montgomery curve \(\EC_{(A,B)}\) is specified by two parameters, 
\(A\) and \(B\).
The most important of the two is \(A\),
which controls the geometry of~\(\EC_{(A,B)}\).
Indeed, the \(j\)-invariant of~\(\EC_{(A,B)}\) is 
\begin{equation}
    \label{eq:j-invariant}
    j(\EC_{(A,B)}) = \frac{256(A^2-3)^3}{A^2 - 4}
    \ ,
\end{equation}
so the \(\FFbar_q\)-isomorphism class of \(\EC_{(A,B)}\)
is completely determined by \(A^2\), 
and independent of \(B\).
We see immediately that not every elliptic curve over \(\FF_q\) 
has a Montgomery model over \(\FF_q\),
since while every element of \(\FF_q\) 
is the \(j\)-invariant of some curve over \(\FF_q\),
not every element of \(\FF_q\) 
is in the form of the right-hand-side of~\eqref{eq:j-invariant}
for some \(A\) in \(\FF_q\)
(we return to the question of which curves have Montgomery models
in~\S\ref{sec:Weierstrass} below).

The parameter \(B\) should be thought of as a ``twisting factor'':
if \(B'\) is another nonzero element of \(\FF_q\),
then \(\EC_{(A,B)}\cong \EC_{(A,B')} \)
via \((x,y) \mapsto (x,\sqrt{B/B'}y)\).
This isomorphism is defined over \(\FF_q\) precisely when \(B/B'\)
is a square in \(\FF_q\).
Otherwise, \(\EC_{(A,B')}\) is a \emph{quadratic twist} of \(\EC_{(A,B)}\):
isomorphic to \(\EC_{(A,B)}\) over \(\FF_{q^2}\), but not over
\(\FF_q\).\footnote{%
    We saw above that since \(j(\EC_{(A,B)})\) is a function of \(A^2\),
    the \(\FFbar_q\)-isomorphism class of \(\EC_{(A,B)}\)
    depends only on \(A^2\).  
    Indeed, \(\EC_{({-A},B)}\)
    is \(\FF_q\)-isomorphic (via \((x,y)\mapsto(-x,y)\))
    to \(\EC_{(A,{-B})}\),
    which is \(\FF_q\)-isomorphic to \(\EC_{(A,B)}\) if \(-1\) is a square in
    \(\FF_q\) (otherwise it is a quadratic twist).
}
All such quadratic twists \(\EC_{(A,B')}\) of \(\EC_{(A,B)}\) 
are isomorphic to each other over
\(\FF_q\): if \(B''\) is another element of \(\FF_q\) 
such that \(B/B''\) is not a square,
then \(B'/B''\) must be a square,
and then \(\EC_{(A,B')}\) and \(\EC_{(A,B'')}\)
are isomorphic over \(\FF_q\) via \((x,y) \mapsto (x,\sqrt{B'/B''} y )\).

The value of \(B\) (modulo squares) 
is more or less incidental for
cryptographic implementations.
In the context of ECM, 
the ability to choose \(B\)
gives us an important degree of freedom for
constructing interesting points in \(\EC(\ZZ/N\ZZ)\).
But when the values of \(B\) and \(B'\) are mathematically and
practically irrelevant (i.e., most of the time),
we will simply call
\(\EC_{(A,B')}\) \emph{the} quadratic twist of \(\EC_{(A,B)}\).
In general, we use \(\EC'\) to denote the quadratic twist of \(\EC\).

\subsection{The group law}
\label{sec:group-law}

Since \(\EC_{(A,B)}\) is an elliptic curve, 
there is a group law \(\oplus\) on its points;
our first task is to describe it.
For the moment, it suffices to work in affine coordinates;
optimized projective formul\ae{} will follow later
in~\S\ref{sec:x-only-arith}.

The point \(\OO\) at infinity acts as the zero element of the group
structure;
the negation map is \(\ominus: (x:y:1) \mapsto (x:-y:1)\).
For addition, 
if \(P = (x_P,y_P)\) and \(Q = (x_Q,y_Q)\) are points on \(\EC_{(A,B)}\),
then \(P \oplus Q = (x_\oplus,y_\oplus)\)
where
\begin{align*}
    x_\oplus & = B\lambda^2 - (x_P+x_Q) - A
    \quad \text{and}
    &
    y_\oplus & = (2x_P+x_Q+A)\lambda-B\lambda^3-y_P
    \\
    & & {} & = \lambda(x_P - x_Q) - y_P
    \ ,
\end{align*}
with
\begin{align*}
    \lambda 
    = 
    \begin{cases}
        (y_Q - y_P)/(x_Q - x_P)
        & \text{ if } P \not= Q \text{ or } {\ominus Q}
        \ ,
        \\
        (3x_P^2+2Ax_P+1)/(2By_P) 
        & \text{ if } P = Q
        \ ;
    \end{cases}
\end{align*}
if \(P = \ominus Q\),
then \(P \oplus Q = \OO\).
We note that \(\lambda\) is the slope of the secant through \(P\)
and \(Q\) (or the tangent to \(\EC_{(A,B)}\) at \(P\), in the case \(P =
Q\)).

As usual, we write \(\EC_{(A,B)}(\FF_q)\) for 
the group of rational points of \(\EC_{(A,B)}\) 
(that is, projective solutions of~\eqref{eq:defining-equation})
with coordinates in \(\FF_q\).
The \(m\)-torsion \(\EC[m]\) is the kernel of the scalar multiplication
\([m]\).
In general, its elements are defined over some extension of \(\FF_q\);
we write \(\EC[m](\FF_q)\) for the group of \(m\)-torsion elements
whose coordinates are in \(\FF_q\).

\subsection{Special torsion and group structures}
\label{sec:torsion}

What can we say about the group structure of \(\EC_{(A,B)}(\FF_q)\)?

We note immediately that \(\EC_{(A,B)}\) always has a rational point of order
two,
\[
    T := (0:0:1) \in \EC_{(A,B)}[2](\FF_q) 
    \ ;
\]
the translation map on \(\EC_{(A,B)}\)
taking \(P\) to \(P \oplus T\)
is 
\begin{equation}
    \label{eq:tau_T}
    \tau_T: (x,y) \longmapsto (1/x,-y/x^2)
    \ .
\end{equation}
In projective coordinates, 
we can see one of the characteristic features of Montgomery curves more
clearly.
On any elliptic curve \(\EC\) with a point \(T\) of order \(2\),
the translation-by-\(T\) map on \(\EC\)
commutes with \(\ominus\),
so it induces an involution on the \(x\)-line
\(\PP^1 = \EC/\subgrp{\ominus}\),
which has the form \((X:Z)\mapsto(aX+bZ:cX+dZ)\).
But for \(T = (0,0)\) on a Montgomery curve \(\EC_{(A,B)}\),
this involution is as simple as possible:
here,
\(\tau_T\) induces \((X:Z) \mapsto (Z:X)\) on~\(\PP^1\).

Montgomery notes in~\cite{montgomery87},
following Suyama,
that the order of \(\EC_{(A,B)}(\FF_q)\)
is always divisible by~4.
The nonsingularity condition requires \(B\), \(A+2\), and \(A-2\) to be
nonzero.
If we let \(\EC_{(A,B')}\) be the quadratic twist of \(\EC_{(A,B)}\),
then the following facts are easy to check:
\begin{enumerate}
    \item
        If \(B(A+2)\) is a square in \(\FF_q\),
        then \((1,\pm\sqrt{(A+2)/B})\) are points of order~4
        in \(\EC_{(A,B)}(\FF_q)\)
        (and if \(B(A+2)\) is not a square, then \((1,\pm\sqrt{(A+2)/B'})\)
        are points of order 4 in \(\EC_{(A,B')}(\FF_q)\)).\footnote{
            Montgomery notes that in an ECM context we can take \(B = A+2\)
            in order to force \((1,1)\) to be a rational point of order 4.
        }
    \item
        Similarly, if \(B(A-2)\) is a square in \(\FF_q\),
        then \((-1,\pm\sqrt{(A-2)/B})\) are points of order~4 
        in \(\EC_{(A,B)}(\FF_q)\);
        otherwise, 
        \((-1,\pm\sqrt{(A-2)/B'})\) are 
        points of order~4 in \(\EC_{(A,B')}(\FF_q)\).
    \item
        If neither \(B(A+2)\) nor \(B(A-2)\) is a square,
        then \(A^2-4\) must be a square;
        then \(x^2 + Ax + 1\) splits completely over \(\FF_p\),
        so \(\EC_{(A,B)}\) has full rational \(2\)-torsion.
        If \(\alpha\) is one root of \(x^2 + Ax + 1\),
        then the other root is \(1/\alpha\);
        the points \((\alpha:0:1)\) and \((1:0:\alpha)\)
        have order~\(2\),
        and are exchanged by~\(\tau_T\).
        There are also points of order two on \(\EC_{(A,B')}\)
        with exactly the same coordinates.
\end{enumerate}
The key thing is that in any finite field,
\(B(A+2)\), \(B(A-2)\), and \(A^2-4\) cannot all be
nonsquares simultaneously: \emph{at least one} of the above situations is
forced to occur, and so \(\#\EC_{(A,B)}(\FF_q)\) is always divisible by \(4\).
Table~\ref{tab:4-torsion} summarizes the resulting group structures 
for various combinations of the three conditions.

\begin{table}
    \centering
    \caption{
        \(4\)-torsion structures on a Montgomery curve \(\EC_{(A,B)}\) 
        and its quadratic twist \(\EC_{(A,B')}\).
    }
    \label{tab:4-torsion}
    \begin{tabular}{c|c||c|c}
        \(B\) & \(A-2\) and \(A+2\) 
        & \(\EC_{(A,B)}[4](\FF_q)\) contains
        & \(\EC_{(A,B')}[4](\FF_q)\) contains
        \\
        \hline
        square & both square 
            & \(\ZZ/4\ZZ\times\ZZ/2\ZZ\)
            & \(\ZZ/2\ZZ\times\ZZ/2\ZZ\)
        \\
        \hline
        square & one square
            & \(\ZZ/4\ZZ\)
            & \(\ZZ/4\ZZ\)
        \\
        \hline
        square & neither square
            & \(\ZZ/2\ZZ\times\ZZ/2\ZZ\)
            & \(\ZZ/4\ZZ\times\ZZ/2\ZZ\)
        \\
        \hline
        nonsquare & both square 
            & \(\ZZ/2\ZZ\times\ZZ/2\ZZ\)
            & \(\ZZ/4\ZZ\times\ZZ/2\ZZ\)
        \\
        \hline
        nonsquare & one square 
            & \(\ZZ/4\ZZ\)
            & \(\ZZ/4\ZZ\)
        \\
        \hline
        nonsquare & neither square
            & \(\ZZ/4\ZZ\times\ZZ/2\ZZ\)
            & \(\ZZ/2\ZZ\times\ZZ/2\ZZ\)
        \\
        \hline
    \end{tabular}
\end{table}

Going further, Suyama shows that if we take
\(A = -(3a^4 + 6a^2 - 1)/4a^3\)
and \(B = (a^2-1)^2/4ab^2\)
for some \(a\) and \(b\) in \(\FF_q\)
with \(ab(a^2-1)(9a^2-1) \not= 0\),
then \((a,b)\) is a point of order \(3\) in \(\EC_{(A,B)}(\FF_q)\).
This is useful in ECM,
where we want to produce curves whose reduction modulo \(p\)
have smooth order; reducing a curve from Suyama's parametrization
yields a curve whose order is divisible by 12, and hence more probably
smooth than a the order of a random curve.

For cryptographic constructions,
we generally want \(\#\EC_{(A,B)}(\FF_q)\) to be as near prime as possible.
A Montgomery curve over a finite field can never have prime order:
we always have \(4\mid\#\EC_{(A,B)}(\FF_q)\)
(and \(4\mid\#\EC_{(A,B')}(\FF_q)\)),
so the best we can hope for is \(\#\EC_{(A,B)}(\FF_q) = 4r\) with \(r\) prime.
Reassuringly, 
there is no theoretical obstruction to 
the existence of \(\EC_{(A,B)}/\FF_q\) 
such that \(\#\EC_{(A,B)}(\FF_q) = 4r\) with \(r\) prime;
and indeed, in practice we have no trouble finding
\(A\) and \(B\) in \(\FF_q\)
such that \(\#\EC_{(A,B)}(\FF_q)/4\) is prime.

\subsection{Correspondence with short Weierstrass models}
\label{sec:Weierstrass}

Any Montgomery curve over \(\FF_q\)
can be transformed into a short Weierstrass model
over \(\FF_q\) (assuming \(q\) is not a power of \(3\)):
for example, the rational maps
\(
    (x,y) \mapsto (u,v) = (B(x + A/3), B^2y)
\)
and \((u,v) \mapsto (x,y) = (u/B-A/3,v/B^2)\)
define an isomorphism over~\(\FF_q\)
between \(\EC_{(A,B)}\) and the short Weierstrass model 
\[
    \EC^W: v^2 = u^3 + (B^2(1-A^2/3)) u + B^3A/3(2{A^2}/{9} - 1)
    \ .
\]

The converse does not hold:
not every short Weierstrass model can be transformed into a Montgomery
model over \(\FF_q\).  
This is obvious enough: not every short Weierstrass model has a
rational point of order \(2\) to map to \(T\)
(and not every \(j\)-invariant in \(\FF_q\) can be expressed 
in the form of~\eqref{eq:j-invariant} with \(A\) in \(\FF_q\)).

Still, it is
useful to have a simple algebraic condition on the coefficients of
a short Weierstrass model that encapsulate its transformability to
Montgomery form.
With this in mind,
Okeya, Kurumatani, and Sakurai~\cite{Okeya--Kurumatani--Sakurai} observe that 
\(\EC^W: v^2 = u^3 + au + b\) has a Montgomery model
if and only if there exist \(\alpha\) and \(\beta\) in \(\FF_q\)
such that 
\(
    \alpha^3 + a\alpha + b = 0
\)
and 
\(
    3\alpha^2 + a = \beta^2
\):
we then have an isomorphism \((u,v) \mapsto (x,y) = ((u - \alpha)/\beta, v/\beta )\)
from \(\EC^W\) to the Montgomery curve \(\EC_{(3\alpha/\beta,1/\beta)}\).
The first relation ensures that 
there is a rational \(2\)-torsion point \((\alpha,0)\) in \(\EC^W(\FF_q)\) 
(which is mapped to \(T = (0,0)\) by the isomorphism);
the second ensures that 
translation by the image of that point 
acts as in~\eqref{eq:tau_T}.

\subsection{Correspondence with twisted Edwards models}
\label{sec:Edwards}

The last decade has seen the great success of Edwards models for
elliptic curves in cryptographic implementations
(see eg.~\cite{edwards07}, \cite{danja07}, and \cite{BBJLP08}). 
It turns out that every Montgomery curve over \(\FF_q\) 
is \(\FF_q\)-isomorphic to a twisted Edwards model,
and vice versa~\cite[\S3]{BBJLP08}.
The rational maps
\begin{align}
    \label{eq:Montgomery-to-Edwards}
    (x,y) \longmapsto & (u,v) = (x/y, (x-1)/(x+1))
    \\
    \label{eq:Edwards-to-Montgomery}
    (u,v) \longmapsto & (x,y) = \left((1+v)/(1-v),(1+v)/((1-v)u\right)
\end{align}
define an isomorphism between \(\EC_{(A,B)}\)
and the twisted Edwards model 
\begin{align}
    \EC_{(a,d)}^\mathrm{Ed}: au^2+v^2 = 1+du^2v^2
    \qquad
    \text{where}
    \quad
    \begin{cases}
        a=(A+2)/B \ ,
        \\
        d=(A-2)/B \ . 
    \end{cases}
\end{align}
The most natural projective closure for \(\EC_{(a,d)}^\mathrm{Ed}\)
is not in \(\PP^2\), 
but in \(\PP^1\times\PP^1\), which embeds into \(\PP^3\)
via the Segre morphism.
Taking coordinates \((U_0:U_1:U_2:U_3)\) on \(\PP^3\),
with \(u = U_1/U_0\), \(v = U_2/U_0\), and \(uv = U_3/U_0\),
we obtain the projective model
\(\EC_{(a,d)}^\mathrm{Ed}: U_0^2 + dU_3^2 = aU_1^2 + U_2^2, U_0U_3 = U_1U_2\)
(the second equation defines the image of \(\PP^1\times\PP^1\)
in~\(\PP^3\));
these are the \emph{extended Edwards coordinates} of~\cite{HWCD}.

The point \(\OO\) on \(\EC_{(A,B)}\)
maps to \((0,1) = (1:0:1:0)\) on \(\EC^\mathrm{Ed}_{(a,d)}\).
The negation is \(\ominus(u,v) = (-u,v)\),
and the map \(\xMap : \EC_{(a,d)}^\mathrm{Ed} \to \PP^1\)
sends \(P\) to \(v(P)\);
if \(\EC_{(a,d)}^\mathrm{Ed}\) is viewed as a curve in \(\PP^1\times\PP^1\),
then \(\xMap\) is just projection onto the second factor.
The distinguished \(2\)-torsion point \(T\) maps to \((0,-1) = (1:0:-1:0)\),
and the translation \(\tau_T\)
becomes \((U_0:U_1:U_2:U_3) \mapsto (U_0:U_1:-U_2:-U_3)\) 
on~\(\EC^\mathrm{Ed}_{(a,d)}\).

We can also consider the images of the other 
torsion points described in~\S\ref{sec:torsion}.
The \(2\)-torsion points \((\frac{-1}{2}(A\pm\sqrt{A^2-4}),0)\)
on \(\EC_{(A,B)}\)
map to \((0:(A-2):0:\pm\sqrt{A^2-4})\) on \(\EC_{(a,d)}^\mathrm{Ed}\),
while 
the 4-torsion points 
\((1,\pm\sqrt{(A+\epsilon2)/B})\)
map to \((\pm\sqrt{(A+\epsilon2)/B}:1:0:0)\) 
for \(\epsilon = \pm 1\).

\section{
    Fast differential arithmetic in $\mathbb{P}^1$
}
\label{sec:x-only-arith}

In projective coordinates,
the quotient map \(\xMap: \EC_{(A,B)}\to\EC/\subgrp{\ominus} = \PP^1\)
is 
\[
    \xMap: P \longmapsto \begin{cases}
        (x_P:1) & \text{ if } P = (x_P:y_P:1) 
        \ ,
        \\
        (1:0)   & \text{ if } P = \OO = (0:1:0) 
        \ .
    \end{cases}
\]
We emphasize that 
the formula \(\xMap((X:Y:Z)) = (X:Z)\)
only holds on the open subset of \(\EC_{(A,B)}\) where \(Z \not= 0\);
it does not extend to the point \(\OO = (0:1:0)\) at infinity,
because \((0:0)\) is not a projective point.

\subsection{Pseudo-operations}
\label{sec:pseudo-ops}

Our first step towards computing
\(\xMap(P)\mapsto\xMap([k]P)\)
is to define efficient pseudo-group operations on \(\PP^1\)
derived from the group operation on \(\EC_{(A,B)}\).
As we noted earlier,
\(\PP^1\) inherits no group structure from \(\EC_{(A,B)}\):
in particular,
there is no map \((\xMap(P),\xMap(Q))\mapsto\xMap(P\oplus Q)\).
This is because \(\xMap(P)\) determines \(P\) only up to sign,
so while \((\xMap(P),\xMap(Q))\) mathematically 
determines the pair \(\{\xMap(P\oplus Q),\xMap(P\ominus Q)\}\),
we cannot tell which of the values is the correct ``sum''.
However,
any three of the values \(\{\xMap(P),\xMap(Q),\xMap(P\oplus
Q),\xMap(P\ominus Q))\}\)
determines the fourth,
so we \emph{can} define a \emph{pseudo-addition} on \(\PP^1\)
by
\[
    \xADD :
    (\xMap(P),\xMap(Q),\xMap(P\ominus Q))
    \longmapsto 
    \xMap(P\oplus Q)
    \ .
\]
The degenerate case where \(P = Q\) becomes a \emph{pseudo-doubling}
\[
    \xDBL :
    \xMap(P) \longmapsto \xMap([2]P)
    \ .
\]
These two operations will be the basis of our efficient
pseudomultiplications.

Our first task is to compute \(\xADD\) and \(\xDBL\) efficiently.
Let \(P = (x_P,y_P)\) and \(Q = (x_Q,y_Q)\) 
be points on \(\EC_{(A,B)}\),
with neither equal to \(T\) or \(\OO\)
(so in particular, \(x_P\) and \(x_Q\) are both nonzero).
Montgomery observed that if \(P \not= Q\),
then the \(x\)-coordinates of \(P\), \(Q\), \(P\oplus Q\), and
\(P\ominus Q\) 
are related by
\begin{equation}
    \label{eq:Montgomery-relation-affine-xADD}
    x_{P\oplus Q}x_{P\ominus Q}(x_P - x_Q)^2 = (x_Px_Q - 1)^2
    \ ,
\end{equation}
while in the case \(P = Q\),
writing \([2]P = (x_{[2]P},y_{[2]P})\),
we have 
\begin{equation}
    \label{eq:Montgomery-relation-affine-xDBL}
    4x_{[2]P}x_P(x_P^2 + Ax_P + 1) = (x_P^2 - 1)^2
    \ .
\end{equation}
Analogous identities exist for general Weierstrass models,
but they are much simpler 
for Montgomery curves---as we will see in~\S\ref{sec:general-x}.
It is this simplicity, due to the special form of \(\tau_T\)
discussed in~\S\ref{sec:torsion},
that leads to particularly efficient pseudo-operations for
Montgomery curves.

\subsection{Pseudo-addition}

Our aim is to compute \(\xMap(P\oplus Q)\)
in terms of \(\xMap(P)\), \(\xMap(Q)\), and \(\xMap(P\ominus Q)\);
we suppose \(P \not= Q\) and \(P\ominus Q \not= T\).
Following Montgomery~\cite[\S10.3.1]{montgomery87},
we move to projective coordinates and write
\begin{align*}
    (X_P:Z_P) 
    & := 
    \xMap(P) 
    \ ,
    &
    (X_Q:Z_Q) 
    & := 
    \xMap(Q) 
    \ ,
    \\
    (X_\oplus:Z_\oplus) 
    & := 
    \xMap(P\oplus Q) 
    \ ,
    &
    (X_\ominus:Z_\ominus)
    & := 
    \xMap(P\ominus Q) 
    \ .
\end{align*}
Since \(P\ominus Q \notin \{\OO,T\}\),
we know that \(X_\ominus \not= 0\) and \(Z_\ominus \not= 0\).
Equation~\eqref{eq:Montgomery-relation-affine-xADD}
therefore becomes the pair of simultaneous relations
\begin{equation}\label{eq:xADDformulas}
    \begin{cases}
        X_{\oplus}
        =
        Z_{\ominus}\left[(X_P-Z_P)(X_Q+Z_Q) + (X_P+Z_P)(X_Q-Z_Q)\right]^2
        \ ,
        \\
        Z_{\oplus}
        =
        X_{\ominus}\left[(X_P-Z_P)(X_Q+Z_Q) - (X_P+Z_P)(X_Q-Z_Q)\right]^2
        \ ,
    \end{cases}
\end{equation}
which Algorithm~\ref{alg:xADD} applies 
to efficiently compute \(\xMap(P\oplus Q)\).
If the ``difference'' \(\xMap(P\ominus Q)\) is fixed 
then we can normalize it to \((X_P:1)\), 
thus saving one multiplication in Step~\ref{alg:xADD:save-here}.
Note that \(\xADD\) involves neither of the curve parameters \(A\) or \(B\),
so it is identical for all Montgomery curves.

\begin{algorithm}
    \caption{\xADD: differential addition on \(\PP^1\)}
    \label{alg:xADD}
    \SetKwInOut{Cost}{Cost}
    \SetKwData{V}{V}
    \KwIn{%
        \((X_P,Z_P)\), \((X_Q,Z_Q)\), and \((X_\ominus,Z_\ominus)\)
        in \(\FF_q^2\) 
        such that \((X_P:Z_P) = \xMap(P)\), \((X_Q:Z_Q) = \xMap(Q)\), and \((X_\ominus:Z_\ominus) = \xMap(P\ominus Q)\)
        for \(P\) and \(Q\) in \(\EC(\FF_q)\) 
    }
    \KwOut{%
        \((X_\oplus,Z_\oplus)\) in \(\FF_q^2\)
        such that
        \((X_\oplus:Z_\oplus) = \xMap(P\oplus Q)\)
        if \(P\ominus Q \notin \{O,T\}\),
        otherwise \(X_\oplus = Z_\oplus = 0\)
    }
    \Cost{\(4\FqM + 2\FqS + 3\Fqa + 3\Fqs\), 
        or \(3\FqM + 2\FqS + 3\Fqa + 3\Fqs\) 
        if \(Z_\ominus\) is normalized to \(1\)}
    \begin{multicols}{2}
        \(\V_0 \gets X_P + Z_P\)
        \tcp{1\Fqa}
        \(\V_1 \gets X_Q - Z_Q\)
        \tcp{1\Fqs}
        \(\V_1 \gets \V_1\cdot \V_0\)
        \tcp{1\FqM}
        \(\V_0 \gets X_P - Z_P\)
        \tcp{1\Fqs}
        \(\V_2 \gets X_Q + Z_Q\)
        \tcp{1\Fqa}
        \(\V_2 \gets \V_2\cdot \V_0\)
        \tcp{1\FqM}
        \(\V_3 \gets \V_1 + \V_2\)
        \tcp{1\Fqa}
        \(\V_3 \gets \V_3^2\)
        \tcp{1\FqS}
        \(\V_4 \gets \V_1 - \V_2\)
        \tcp{1\Fqs}
        \(\V_4 \gets \V_4^2\)
        \tcp{1\FqS}
        \(X_\oplus \gets Z_\ominus\cdot \V_3\)
        \label{alg:xADD:save-here}
        \tcp{\(1\FqM\) / \(0\FqM\) if \(Z_\ominus = 1\)}
        \(Z_\oplus \gets X_\ominus\cdot \V_4\)
        \tcp{1\FqM}
        \Return{\((X_\oplus:Z_\oplus)\)}
    \end{multicols}
\end{algorithm}

\subsection{Pseudo-doubling}\label{sub:pseudo-doubling}

It remains to handle the doubling case, where \(Q = P\). 
We want to compute
\(
    (X_{[2]P}:Z_{[2]P})
    := 
    \xMap([2]P)
\)
in terms of \(\xMap(P)\).
Again, we follow~\cite[\S10.3.1]{montgomery87}:
in projective coordinates,
Equation~\eqref{eq:Montgomery-relation-affine-xDBL}
becomes the pair of simultaneous relations
\begin{equation}\label{eq:xDBLformulas}
    \begin{cases}
        X_{[2]P}
        =  
        (X_P + Z_P)^2(X_P - Z_P)^2
        \ ,
        \\
        Z_{[2]P}
        = 
        (4X_PZ_P)((X_P-Z_P)^2 + ((A+2)/4)(4X_PZ_P))
        \ .
    \end{cases}
\end{equation}
Algorithm~\ref{alg:xDBL} 
uses these, and the identity
\(
    4X_PZ_P = (X_P + Z_P)^2 - (X_P - Z_P)^2
\),
to efficiently compute \((X_{[2]P}:Z_{[2]P})\).
The fact that the right-hand-sides of~\eqref{eq:xDBLformulas}
are symmetric under \((X_P:Z_P)\leftrightarrow(Z_P:X_P)\)
reflects the fact that the doubling map \([2]\) 
factors through the \(2\)-isogeny
\(\EC_{(A,B)} \to \EC_{(A,B)}/\subgrp{T}\);
this aspect of Montgomery's \(x\)-line arithmetic 
later found an echo in Doche, Icart, and Kohel's work on efficient doubling
and tripling~\cite{DIK}.

\begin{algorithm}
    \caption{\xDBL: pseudo-doubling on \(\PP^1\) from
    \(\EC_{(A,B)}\)}
    \label{alg:xDBL}
    \SetKwInOut{Cost}{Cost}
    \SetKwData{V}{V}
    \KwIn{%
        \((X_P,Z_P)\) in \(\FF_q^2\) 
        such that \((X_P:Z_P) = \xMap(P)\) for \(P\) in \(\EC(\FF_q)\)
    }
    \KwOut{%
        \((X_{[2]P},Z_{[2]P})\) in \(\FF_q^2\)
        such that \((X_{[2]P}:Z_{[2]P}) = \xMap([2]P)\)
        if \(P \notin \{O,T\}\), 
        otherwise \(Z_{[2]P} = 0\)
    }
    \Cost{\(2\FqM + 2\FqS + 1\FqMconst + 3\Fqa + 1\Fqs\)}
    \begin{multicols}{2}
        \BlankLine
        \(\V_1 \gets X_P + Z_P\)
        \tcp{1\Fqa}
        \(\V_1 \gets \V_1^2\)
        \tcp{1\FqS}
        \(\V_2 \gets X_P - Z_P\)
        \tcp{1\Fqs}
        \(\V_2 \gets \V_2^2\)
        \tcp{1\FqS}
        \(X_{[2]P} \gets \V_1\cdot\V_2\)
        \tcp{1\FqM}
        \(\V_1 \gets \V_1 - \V_2\)
        \tcp{1\Fqs}
        \(\V_3 \gets ((A+2)/4)\cdot\V_1\)
        \tcp{1\FqMconst}
        \label{alg:xDBL:const}
        \(\V_3 \gets \V_3 + \V_2\)
        \tcp{1\Fqa}
        \(Z_{[2]P} \gets \V_1\cdot\V_3\)
        \tcp{1\FqM}
        \Return{\((X_{[2]P}:Z_{[2]P})\)}
    \end{multicols}
\end{algorithm}

Step~\ref{alg:xDBL:const} of Algorithm~\ref{alg:xDBL}
is a multiplication by the constant \((A + 2)/4\).
We emphasize that this is the \emph{only} place in the \(x\)-line arithmetic
where the parameter~\(A\) appears. 
The parameter~\(B\) never appears at all: 
\(x\)-line arithmetic is twist-agnostic.
For implementations, therefore,
we try to choose \(A\)
such that the cost of multiplying by \((A+2)/4\) is minimised 
(for example, taking \(A\) such that \((A+2)/4\) is particularly small).

\subsection{Comparison with general \(x\)-line arithmetic}
\label{sec:general-x}

To see how Montgomery's choice of curve model 
contributes to efficient \(x\)-line arithmetic, 
it is instructive to compare the pseudo-addition
with the equivalent formul\ae{} 
for a general Weierstrass model
\[
    \EC: y^2 = x^3 + f_2x^2 + f_1x + f_0
    \ .
\]
The analogues of~\eqref{eq:Montgomery-relation-affine-xADD} 
and~\eqref{eq:Montgomery-relation-affine-xDBL}
for \(\EC\)
are 
\(
    x_\oplus x_\ominus (x_P - x_Q)^2
    =
    (x_Px_Q - f_1)^2 - 4f_0(x_P + x_Q + f_2) 
\)
and
\(
    4x_{[2]P}( x_P^3 + f_2x_P^2 + f_1x_P + f_0 )
    =
    (x_P^2 - f_1)^2 - 4f_0(2x_P + f_2)
\),
respectively.
In projective coordinates, these become 
the relatively complicated pseudo-addition formul\ae{}
\[
    \begin{cases}
        X_\oplus 
        = 
        Z_\ominus\big[
            (X_PX_Q - f_1Z_PZ_Q)^2
            - 4f_0(Z_PX_Q + X_PZ_Q + f_2Z_PZ_Q)Z_PZ_Q
        \big]
        \\
        Z_\oplus 
        = 
        X_\ominus (Z_PX_Q - X_PZ_Q)^2
    \end{cases}
    \!\!\!\!\!\!\!\!
\]
and pseudo-doubling formul\ae{} 
\[
    \begin{cases}
        X_{[2]P} 
        = 
        (X_P^2 - f_1Z_P^2)^2 - 4f_0(2X_PZ_P + f_2Z_P^2)Z_P^2
        \ ,
        \\
        Z_{[2]P}
        =
        4(X_P^4 + f_2X_P^2Z_P^2 + f_1X_PZ_P^3 + f_0Z_P^4)
        \ .
    \end{cases}
\]

If we specialize and take \(f_2 = 0\), 
leaving \(f_1\) and \(f_0\) free,
then we are in the case of short Weierstrass models
(for which the affine formul\ae{} are classical:
see eg.~\cite[Formulary]{Cassels}).
This imposes no special structure on \(\EC\),
since every elliptic curve is isomorphic to
a short Weierstrass model over \(\FF_q\).
The resulting projective formul\ae{} were proposed for
side-channel-aware implementations by Brier and
Joye~\cite[\S4]{Brier--Joye},
with pseudo-addition requiring \(7\FqM + 2\FqS + 2\FqMconst + 3\Fqa + 2\Fqs\)
and pseudo-doubling \(3\FqM+4\FqS+2\FqMconst+6\Fqa+2\Fqs\).
     
But we can simplify things more dramatically by taking \(f_0 = 0\) instead, 
leaving \(f_1\) and \(f_2\) free.
This is equivalent to requiring a rational \(2\)-torsion point on \(\EC\),
which we move to \((0,0)\).
The pseudo-addition formul\ae{} become
\[
    ( X_\oplus : Z_\oplus )
    = 
    \left(
        Z_\ominus(X_PX_Q - f_1Z_PZ_Q)^2
        :
        X_\ominus (Z_PX_Q - X_PZ_Q)^2 
    \right)
    \ ,
\]
which can be evaluated in \(6\FqM + 2\FqS + 1\FqMconst + 2\Fqs\),
while pseudo-doubling becomes
\[
    ( X_{[2]P} : Z_{[2]P} )
    = 
    \left( 
        (X_P^2 - f_1Z_P^2)^2 
        :
        4(X_P^4 + f_2X_P^2Z_P^2 + f_1X_PZ_P^3)
    \right)
    \ ,
\]
which can be evaluated in 
\(2\FqM + 4\FqS + 2\FqMconst + 4\Fqa + 1\Fqs\).

Now taking \(f_1 = 1\) 
not only eliminates \(1\FqMconst\) in the pseudo-addition
and pseudo-doubling,
it also allows us to save \(2\FqM\) in the pseudo-addition
and \(2\FqS\) in the pseudo-doubling
(at the cost of a few more additions and subtractions)
by exploiting the symmetry of the resulting forms\footnote{%
    Translation by the \(2\)-torsion point \((0,0)\)
    is defined by \((x,y) \mapsto (f_1/x,-f_1y/x^2)\);
    taking \(f_1 = 1\)
    is therefore equivalent to putting this translation map
    in the special form of~\eqref{eq:tau_T}.
}.
Indeed,
if we write \(A\) for \(f_2\) and allow a possible quadratic twist by \(B\),
then we have arrived at Montgomery's model
\(\EC: By^2 = x(x^2 + Ax + 1)\),
and we recover the efficient pseudo-addition and pseudo-doubling 
in~\eqref{eq:xADDformulas}
and~\eqref{eq:xDBLformulas}.

\section{
    The Montgomery ladder
}
\label{sec:ladder}

We now resume our task of computing 
\(
    \xMap(P) \mapsto \xMap([k]P)
\).
We begin by explaining a version of the Montgomery ladder that uses full
group operations to compute \(P \mapsto [k]P\) on \(\EC_{(A,B)}\), 
before deriving the classic \(x\)-line Montgomery ladder
which computes \(\xMap(P) \mapsto \xMap([k]P)\)
using only \(\xADD\) and \(\xDBL\).
We then present Okeya and Sakurai's version of the L\'opez--Dahab
trick, an appendix to the \(x\)-only ladder
which recovers the full image point \([k]P\) on \(\EC_{(A,B)}\).

\subsection{The ladder in a group}

Algorithm~\ref{alg:ladder-group} presents the Montgomery ladder
algorithm in the context of a group, for ease of analysis.
(While the algorithm is presented using a Montgomery curve
\(\EC_{(A,B)}\), it is clear that it works in any abelian group).

\begin{algorithm}
    \caption{Montgomery's binary algorithm in the 
    group \(\EC_{(A,B)}(\FF_q)\)}
    \label{alg:ladder-group}
    \SetKwInOut{Cost}{Cost}
    \SetKwData{R}{R}
    \KwIn{%
        \(k = \sum_{i=0}^{\ell-1}k_i2^i\) with \(k_{\ell-1} = 1\),
        and \(P \in \EC(\FF_q)\)
    }
    \KwOut{$[k]P$}
    \Cost{\(\ell-1\) calls to \(\oplus\) and \(\ell\) calls to \([2]\)}
    $(\R_0,\R_1) \leftarrow (P,[2]P)$
    \label{alg:ladder-group:init}
    \;
    \For{$i=\ell-2$ {\bf down to} $0$}{
        \uIf{$k_i = 0$}{
            $(\R_0,\R_1) \leftarrow ([2]\R_0,\R_0\oplus\R_1)$
            \label{alg:ladder-group:zero-bit}
        }
        \Else{
            $(\R_0,\R_1) \leftarrow (\R_0\oplus\R_1,[2]\R_1)$
            \label{alg:ladder-group:one-bit}
        }
    }
    \Return{ \(\R_0\) }
    \label{alg:ladder-group:end}
\end{algorithm}

Algorithm~\ref{alg:ladder-group} maintains two important invariants.
First, looking at 
Lines~\ref{alg:ladder-group:zero-bit}
and~\ref{alg:ladder-group:one-bit},
we see that the difference \(R_1\ominus R_0\) never changes;
then, looking at Line~\ref{alg:ladder-group:init},
we see that that difference must always be \(P\);
hence, at all times, 
\begin{align}
    \label{eq:Monty-invariant-1}
    R_1 = R_0 \oplus P
    \ .
\end{align}
Second, 
after iteration \(i\) of the loop (counting downwards from \(\ell-2\)),
we have 
\begin{align}
    \label{eq:Monty-invariant-2}
    R_0 = [(k)_i]P
    \quad
    \text{and}
    \quad
    R_1 = [(k)_i+1]P
    \ ,
    \quad
    \text{where}
    \quad
    (k)_i := \lfloor{k/2^i}\rfloor
    \ .
\end{align}
This proves the correctness of Algorithm~\ref{alg:ladder-group}:
at Line~\ref{alg:ladder-group:end} we have just finished iteration \(i = 0\),
so we return \(R_0 = [(k)_0]P = [k]P\).
Equation~\eqref{eq:Monty-invariant-2}
is easy to see once we know~\eqref{eq:Monty-invariant-1}:
looking at \(R_0\) in each iteration,
we recover the classic double-and-add method for computing \([k]P\).
First, \(R_0\) is initialized to~\(P\);
then, 
if \(k_i = 0\) we double \(R_0\),
while if \(k_i = 1\) we replace \(R_0\) with \(R_0\oplus R_1\),
which is \([2]R_0\oplus P\) by~\eqref{eq:Monty-invariant-1}---that
is, we double \(R_0\) and add \(P\).

\subsection{The Montgomery ladder}

Equations~\eqref{eq:Monty-invariant-1} and~\eqref{eq:Monty-invariant-2}
allow us to transform Algorithm~\ref{alg:ladder-group} 
into Algorithm~\ref{alg:ladder},
known as the Montgomery ladder.
We want to compute 
\(
    \xMap(P) \mapsto \xMap([k]P) 
\).
We may suppose \(P\not= O\),
since \(\xMap([k]O) = \xMap(O)\) for all \(k\),
and we may also suppose \(P \not= T\),
since \(\xMap([k]T) = \xMap(T)\) for odd \(k\)
and \(\xMap(O)\) for even \(k\).

Maintaining the notation of~\eqref{eq:Monty-invariant-2},
let \SetKwData{x}{x}\((\x_0,\x_1) = (\xMap([(k)_i]P),\xMap([(k)_i+1]P))\).
Then~\eqref{eq:Monty-invariant-1}
shows that we can compute
\((\xMap([(k)_{i-1}]P),\xMap([(k)_{i-1}+1]P))\)
as
\((\xDBL(\x_0),\xADD(\x_0,\x_1,\xMap(P)))\)
or \((\xADD(\x_0,\x_1,\xMap(P)),\xDBL(\x_1))\),
depending on the value of the bit \(k_i\).\footnote{%
    Since the \(\xADD\) and \(\xDBL\) calls always share an
    argument, it is common for high-performance implementations
    to exploit any overlap between intermediate calculations 
    in the \(\xADD\) and \(\xDBL\)
    by merging them in one combined function. 
}
We can therefore
initialize \((\x_0,\x_1)\) 
to \((\xMap(P),\xMap([2]P)) = (\xMap(P),\xDBL(\xMap(P)))\),
and then applying the transitions above for each bit of~\(k\)
will yield
\((\x_0,\x_1) = (\xMap([k]P),\xMap([k+1]P))\). 

While the final value of \(\x_0\) is \(\xMap([k]P)\),
the target of our calculation,
we will see in~\S\ref{sec:recovery}
that the final value \(\xMap([k+1]P)\) of \(\x_1\)
can be used to help recover the full group element \([k]P\),
if desired.
We therefore include \(\x_1\) as an optional second return value
in Algorithm~\ref{alg:ladder}.

\begin{algorithm}
    \caption{\LADDER: The Montgomery ladder}
    \label{alg:ladder}
    \SetKwInOut{Cost}{Cost}
    \SetKwData{x}{x}
    \KwIn{%
        \(k = \sum_{i=0}^{\ell-1}k_i2^i\) with \(k_{\ell-1}=1\), 
        and \((X_P,Z_P)\) in \(\FF_q^2\) s.t. \((X_P:Z_P) = \xMap(P)\)
    }
    \KwOut{%
        \((X_k,Z_k) \in \FF_q^2\) 
        s.t. \((X_k:Z_k) = \xMap([k]P)\) if \(P \notin \{O,T\}\),
        otherwise \(Z_k = 0\).
        Also optionally returns \((X_{k+1},Z_{k+1}) \in \FF_q^2\)
        s.t. \((X_{k+1}:Z_{k+1}) = \xMap([k]P)\) if \(P \notin \{O,T\}\),
        otherwise \(Z_{k+1} = 0\).
    }
    \Cost{\(\ell-1\) calls to \xADD\ and \(\ell\) calls to \xDBL}
    \(
        (\x_0,\x_1) 
        \leftarrow 
        ((X_P,Z_P),\xDBL((X_P,Z_P)))
    \)
    \label{eq:Monty-initialize}
    \;
    \For{$i=\ell-2$ {\bf down to} $0$}{
        \uIf{$k_i = 0$}{
            \(
                (\x_0,\x_1) 
                \gets 
                \left(
                    \xDBL(\x_0) ,
                    \xADD(\x_0,\x_1,(X_P,Z_P)) 
                \right)
            \)
        }
        \Else{
            \(
                (\x_0,\x_1) 
                \gets 
                \left(
                    \xADD(\x_0,\x_1,(X_P,Z_P)) ,
                    \xDBL(\x_1)
                \right)
            \)
        }
    }
    \Return{ \(\x_0\) (and optionally \(\x_1\)) } \label{line:ladder-return}
\end{algorithm}

\subsection{Recovery of $y$-coordinates}
\label{sec:recovery}

On the surface,
the Montgomery ladder 
appears to be an algorithm for computing \(\xMap([k]P)\) 
from \(\xMap(P)\).
However, L\'opez and Dahab~\cite{LD99} observed that since 
it actually computes \(\xMap([k]P)\) and \(\xMap([k+1]P)\),
the ladder can easily be extended to compute the full scalar multiplication
\(P\mapsto [k]P\)
by first computing
\(\xMap(P)\mapsto(\xMap([k]P),\xMap([k+1]P))\)
and then recovering \([k]P\)
from the data \((P,\xMap([k]P),\xMap([k+1]P))\).

L\'opez and Dahab originally gave formul\ae{} for this recovery step
specific to Montgomery curves over binary fields.
Their results were extended to prime-field Montgomery curves
by Okeya and Sakurai~\cite{Okeya--Sakurai},
and later to short Weierstrass models by Brier and Joye~\cite{Brier--Joye}.
Kohel provides a more general and powerful point of view in~\cite{Kohel11},
treating the image of
the curve under the map \(Q \mapsto (\xMap(Q),\xMap(Q\oplus P))\)
as a new model of the elliptic curve itself.

Okeya and Sakurai proceed as follows.
Suppose \(P\) is not in \(\EC_{(A,B)}[2]\),
and \(Q\) is not in \(\{P,{\ominus P},\OO\}\).\footnote{%
    This is not a serious restriction in the context of
    scalar multiplication, where \(Q = [k]P\):
    if \(P\) is a point of order \(2\),
    then either $[k]P = O$ or \(y([k]P) = 0\).
}
In affine coordinates, 
writing
\( (x_P,y_P) = P \),
\( (x_Q,y_Q) = Q \),
and
\( (x_\oplus,y_\oplus) = P \oplus Q \)
as usual,
the group law formul\ae{} in \S\ref{sec:group-law} show that
\(y_Q\) can be deduced from \(x_P\), \(y_P\), \(x_Q\), and \(x_\oplus\) 
using the relation
\[
    y_Q
    =
    \frac{(x_Px_Q+1)(x_P+x_Q+2A)-2A-(x_P - x_Q)^2x_\oplus}{2By_P}
\]
(note the re-appearance of the twisting parameter \(B\)).

Algorithm~\ref{alg:Recover},
taken from~\cite[Algorithm~1]{Okeya--Sakurai},
applies this to compute \(Q\) from \(P\), \(\xMap(Q)\), 
and \(\xMap(P\oplus Q)\).
Lines~\ref{alg:recover:constant1}
and~\ref{alg:recover:constant2} 
involve multiplications by the constants $2A$ and $2B$. 
Referring
back to~\S\ref{sub:pseudo-doubling}, if $(A+2)/4$ is chosen to be
advantageously small, 
then \(2A = 8((A+2)/4)-4\) is also small. Similarly, referring back
to~\S\ref{sub:paramsAB}, we can choose $B$ such that multiplication by \(2B\) is essentially free. 

\begin{algorithm}
    \caption{\Recover: Okeya--Sakurai \(y\)-coordinate recovery}
    \label{alg:Recover}
    \SetKwInOut{Cost}{Cost}
    \SetKwData{V}{v}
    \KwIn{\((x_P:y_P:1) = P\), \((X_Q:Z_Q) = \xMap(Q)\), 
        and \((X_\oplus:Z_\oplus) = \xMap(P\oplus Q)\)
        for \(P\) and \(Q\) in \(\EC_{(A,B)}(\FF_q)\) with \(P \notin
        \EC_{(A,B)}[2]\) and \(Q \notin \{P,\ominus P,O\}\).
    }
    \KwOut{\((X':Y':Z') = Q\)}
    \Cost{\(10\FqM + 1\FqS + 2\FqMconst + 3\Fqa + 3\Fqs\)}
    \begin{multicols}{3}
        \(\V_1 \gets x_P\cdot Z_Q\)
        \tcp{1\FqM}
        \(\V_2 \gets X_Q + \V_1\)
        \tcp{1\Fqa}
        \(\V_3 \gets X_Q - \V_1\)
        \tcp{1\Fqs}
        \(\V_3 \gets \V_3^2\)
        \tcp{1\FqS}
        \(\V_3 \gets \V_3\cdot X_\oplus\)
        \tcp{1\FqM}
        \(\V_1 \gets 2A \cdot Z_Q\) \label{alg:recover:constant1}
        \tcp{1\FqMconst}
        \(\V_2 \gets \V_2 + \V_1\)
        \tcp{1\Fqa}
        \(\V_4 \gets x_P\cdot X_Q\)
        \tcp{1\FqM}
        \(\V_4 \gets \V_4 + Z_Q\)
        \tcp{1\Fqa}
        \(\V_2 \gets \V_2 \cdot \V_4\)
        \tcp{1\FqM}
        \(\V_1 \gets \V_1 \cdot Z_Q\)
        \tcp{1\FqM}
        \(\V_2 \gets \V_2 - \V_1\)
        \tcp{1\Fqs}
        \(\V_2 \gets \V_2\cdot Z_\oplus\)
        \tcp{1\FqM}
        \(Y' \gets \V_2 - \V_3\)
        \tcp{1\Fqs}
        \(\V_1 \gets 2B \cdot y_P\) \label{alg:recover:constant2}
        \tcp{1\FqMconst}
        \(\V_1 \gets \V_1 \cdot Z_Q\)
        \tcp{1\FqM}
        \(\V_1 \gets \V_1 \cdot Z_\oplus\)
        \tcp{1\FqM}
        \(X' \gets \V_1\cdot X_Q\)
        \tcp{1\FqM}
        \(Z' \gets \V_1\cdot Z_Q\)
        \tcp{1\FqM}
        \Return{\((X':Y':Z')\)}
    \end{multicols}
\end{algorithm}

Combining Algorithms~\ref{alg:ladder} and~\ref{alg:Recover}
yields Algorithm~\ref{alg:ladder-plus-Recover},
an efficient scalar multiplication routine for the full group~\(\EC_{(A,B)}(\FF_q)\)
using \(x\)-only arithmetic.
This is generally much more efficient than Algorithm~\ref{alg:ladder-group}.

\begin{algorithm}
    \caption{Scalar multiplication on \(\EC_{(A,B)}(\FF_q)\),
        combining the Montgomery ladder with \(y\)-coordinate recovery}
    \label{alg:ladder-plus-Recover}
    \SetKwInOut{Cost}{Cost}
    \SetKwData{V}{x}
    \KwIn{%
        $k \in \ZZ_{>0}$, 
        and $P$ in \(\EC_{(A,B)}(\FF_q)\setminus\EC_{(A,B)}[2](\FF_q)\)
    }
    \KwOut{\([k]P\)}
    %
    \((\V_0,\V_1) \gets \LADDER(k,(X_P,Z_P))\)
    where \((X_P:Z_P) = \xMap(P)\)
    \;
    \(Q \gets \Recover(P,\V_0,\V_1)\)
    \;
    \Return{\(Q\)}
\end{algorithm}

\section{
    Montgomery curves and ladders in elliptic curve cryptography 
}
\label{sec:ECC}


We saw in~\S\ref{sec:torsion} that no Montgomery curve can have prime order,
since their order is always divisible by 4.
The presence of this small cofactor 4 has no serious impact on the
security level of a well-chosen Montgomery curve, since the state of the art for solving discrete logarithms in large
prime-order subgroups of Montgomery curves is still Pollard's rho method~\cite{pollardrho}.

\subsection{Montgomery curves in cryptographic standards}
\label{sec:cryptographic-params}

None of the elliptic curves so far standardardized by
NIST~\cite{FIPS186-4}, Brainpool~\cite{brainpool}, or ANSSI~\cite{ANSSI}
can be transformed into Montgomery form over the base field,
because they all have prime order, so they are
unfortunately incompatible with Montgomery arithmetic. However, the two curves recently proposed by the Internet Research Task Force (IRTF) for standardization in the Transport Layer Security (TLS) protocol are specified in Montgomery form~\cite{RFC}.

\begin{example}[Curve25519~\cite{curve25519}]\label{eg:curve25519}
    The most widely-known Montgomery curve in contemporary
    cryptography is the curve used in Bernstein's Curve25519
    software for Diffie--Hellman key exchange.
    This curve is defined by
    \[
        \EC/\FF_p: y^2 = x(x^2 + 486662x + 1)
        \quad \text{where} \quad
        p = 2^{255} - 19
        \ .
    \]
    We find \(\#\EC(\FF_p) = 8r\) 
    and \(\#\EC'(\FF_p) = 4r'\), where $r$ and $r'$
    are 253-bit primes. 
\end{example}

\begin{example}[Curve448~\cite{goldilocks}]
    Hamburg's Curve448 offers 
    a conservative, high-strength alternative to Curve25519 for TLS. 
    This curve is defined by
    \[
        \EC/\FF_p: y^2 = x(x^2 + 156326x + 1)
        \quad \text{where} \quad
        p = 2^{448} - 2^{224}-1
        \ .
    \]
    We find \(\#\EC(\FF_p) = 4r\) 
    and \(\#\EC'(\FF_p) = 4r'\), where $r$ and $r'$
    are 446- and 447-bit primes, respectively. 
\end{example}

\subsection{Diffie--Hellman with $x$-coordinates}
\label{sec:x-only-DH}

Recall Miller's \(x\)-only Diffie--Hellman protocol, described in~\S\ref{sec:x-only-DH-intro}:
Alice computes $(a,\xMap(P)) \mapsto \xMap([a]P)$ and transmits her public key $\xMap([a]P)$ to Bob. Upon receiving Bob's public key $\xMap([b]P)$, she computes $(a,\xMap([b]P)) \mapsto \xMap([ab]P)$ to arrive at the same shared secret as Bob, who computes $(b,\xMap([a]P)) \mapsto \xMap([ba]P)$. 
This \(x\)-only protocol is the basis of Bernstein's Curve25519 key exchange software. 
But in addition to working entirely with $x$-coordinates, Bernstein
observed that the Montgomery form allows for another simplification in
real-world implementations. Recall from~\S\ref{sub:paramsAB} that if
$\EC_{(A,B')}(\FF_q)$ is the quadratic twist of
$\EC_{(A,B)}(\FF_q)$, then precisely one of $B$ and $B'$ is a square in
$\FF_q$, and the other is a non-square. It follows that every
element $x_P$ in $\FF_q$ corresponds to a point $P=(x_P,y_P)$ which is
either in $\EC_{(A,B)}(\FF_q)$, in $\EC_{(A,B')}(\FF_q)$, or (if
$y_P=0$) in both \(\EC_{(A,B)}[2](\FF_q)\) and \(\EC_{(A,B')}[2](\FF_q)\).
Bernstein chose the curve in Example~\ref{eg:curve25519} by insisting
that both $\EC_{(A,B)}(\FF_q)$ and $\EC_{(A,B')}(\FF_q)$ have
cryptographically strong (i.e., almost prime) group orders. This means
that every element of $\FF_q$ corresponds to a point on a
cryptographically strong Montgomery curve, and implementers need not
perform any \emph{point validation} checks~\cite{invalidcurve}
in this protocol.
Moreover, since the Montgomery ladder does not use the constant $B$, it
correctly computes $(\xMap(P),k) \mapsto \xMap([k]P)$ irrespective of
the twist that $P$ lies on. If both $\EC_{(A,B)}$ and $\EC_{(A,B')}$ are
secure, then $\EC_{(A,B)}$ is said to be \emph{twist-secure}.

As we noted above,
when implementing cryptosystems based on 
the discrete logarithm or Diffie--Hellman problems,
we generally want \(\#\EC_{(A,B)}(\FF_q)\) 
as close to prime as possible.
We have \(\#\EC_{(A,B)}(\FF_q) + \#\EC_{(A,B')}(\FF_q) = 2q + 2\),
as with any elliptic curve-twist pair over \(\FF_q\);
so if \(\EC_{(A,B)}(\FF_q) = 4r\) with \(r\) prime,
then the closest we can come to prime order for the twist is
\(\#\EC_{(A,B')}(\FF_q) = 4r'\) with \(r'\) prime
if \(q \equiv 3 \pmod{4}\),
and \(\#\EC_{(A,B')}(\FF_q) = 8r'\) with \(r'\) prime
if \(q \equiv 1 \pmod{4}\).
We do not know of any theoretical asymptotic results
guaranteeing a density or distribution 
of twist-secure Montgomery curves over any finite field,
but there does not seem to be
any problem in finding such curves in practice.\footnote{%
    An analysis of the frequency of prime-order curves with prime-order twists
    appears in~\cite{Shparlinski--Sutantyo}; but this does not
    apply to Montgomery curves, since they cannot have prime order.
}

The presence of non-trivial cofactors means that care must be taken in
certain scenarios to thwart the threat of small subgroup attacks~\cite{limlee}.
For $x$-line Diffie--Hellman, the easiest way to do this is to define all secret scalars to be a multiple of the lowest common multiple of the curve and twist cofactors.

\subsection{Constant-time ladders}

For secure software implementations of ECC,
it is important that scalar multiplication routines exhibit uniform execution patterns with no correlation between timing and secret data; such \emph{constant-time} behaviour is an essential first step towards preventing timing attacks~\cite{Kocher}. 
Unlike many other addition chains and scalar multiplication algorithms,
the Montgomery ladder has an inherently uniform execution pattern.
Nevertheless, a number of issues must still be addressed in order to
achieve constant-time implementations. 

As it stands, the length of the main loop in Algorithm~\ref{alg:ladder}
is determined by the bitlength \(k\) of the input scalar $m \in [0,r)$.
There are two common strategies for making the loop length independent of $k$.
One option is to require all scalars to have their top bit set, either
by defining them that way (as was done in~\cite{curve25519}) or by
adding a small, fixed multiple of the (sub)group order to each scalar.
A second option is to modify Step~\ref{eq:Monty-initialize} of Algorithm~\ref{alg:ladder},
setting \SetKwData{x}{x}$(\x_0,\x_1) \leftarrow (\xMap(O),\xMap(P))$
instead of $(\xMap(P),\xMap([2]P))$.
Since the formul\ae{} for $\xDBL$ and $\xADD$ behave correctly under these inputs, $\x_0$ and $\x_1$ will remain unchanged until the first non-zero bit of the scalar $k$ is encountered, so a constant-length loop can be achieved by accepting scalars as all bitstrings of a fixed length. 

Algorithm~\ref{alg:ladder}
presents the ladder using an \textbf{if} statement.
Since each \textbf{if} represents branching on potentially secret data,
and these branches may be measured in timing variations,
it is standard practice to replace the branches with conditional swaps
(such as the \(\CSWAP\) defined in Algorithm~\ref{alg:CSWAP})
to avoid leaking information on secret scalars.
The result is Algorithm~\ref{alg:uniform-monty},
which consists of a uniform sequence of \(\xDBL\)s and \(\xADD\)s.
Provided the field arithmetic used by the \(\xADD\) and \(\xDBL\) 
specified in Algorithms~\ref{alg:xADD} and~\ref{alg:xDBL}
is implemented in a completely uniform way,
this yields a completely uniform algorithm.

\begin{algorithm}
    \caption{\CSWAP: Constant-time conditional swap.}
    \label{alg:CSWAP}
    \SetKwData{x}{x}
    \KwIn{\(b \in \{0,1\}\) and a pair \((\x_0,\x_1)\) 
        of objects encoded as \(n\)-bit strings%
    }
    \KwOut{\((\x_b,\x_{1-b})\)}
    \SetKwData{V}{v}
    \SetKwData{b}{b}
    \(\b \gets (b,\ldots,b)_n\)
    \;
    \(\V \gets \b\ \mathtt{and}\ (\x_0\ \mathtt{xor}\ \x_1)\)
    \tcp*{bitwise and, xor; do not short-circuit and}
    \Return{\((\x_0\ \mathtt{xor}\ \V, \x_1\ \mathtt{xor}\ \V)\)}
    \;
\end{algorithm}

\begin{algorithm}
    \caption{A uniform Montgomery ladder}
    \label{alg:uniform-monty}
    \SetKwInOut{Cost}{Cost}
    \SetKwData{t}{x}
    \KwIn{\(k = \sum_{i=0}^{\ell-1}k_i2^i\)
        with \(k_{\ell-1} =1\), and \(\xMap(P)\) for \(P\) in
        \(\EC_{(A,B)}(\FF_q)\)
    }
    \KwOut{%
        \((X_k,Z_k) \in \FF_q^2\) 
        s.t. \((X_k:Z_k) = \xMap([k]P)\) if \(P \notin \{O,T\}\),
        otherwise \(Z_k = 0\).
    }
    \Cost{ \(\ell-1\) calls to \(\xADD\), \(\ell\) calls to \(\xDBL\), and
    \(\ell-1\) calls to \(\CSWAP\) }
    \( (\t_0,\t_1) \gets (\xDBL((X_P,Z_P)),(X_P,Z_P)) \) 
    \;
    \For{$i = \ell-2$ {\bf down to} $0$}{
        \((\t_0,\t_1) \gets \CSWAP((k_{i+1}\ \mathtt{xor}\ k_i),(\t_0,\t_1)) \) 
        \;
        \((\t_0,\t_1) \gets (\xDBL(\t_0),\xADD(\t_0,\t_1,(X_P,Z_P))) \) 
        \;
    }
    \((\t_0,\t_1) \gets \CSWAP(k_0,(\t_0,\t_1)) \) 
    \;
    \Return{\(\t_0\)}
    \;
\end{algorithm}

\subsection{Completeness and Bernstein's modified \(\xMap\)-map}

Algorithms~\ref{alg:ladder} and~\ref{alg:uniform-monty}
do not compute the pseudomultiplication \(\xMap([k]P)\) correctly
if \(P = O\) or \(T\):
instead, they return \((X_k,Z_k) = (e,0)\) for some \(e\) in \(\FF_q\)
(see~\cite[Theorem 4.3]{avoidzero} for a more precise statement).
In some cases \(e\) may be \(0\),
in which case \((X_k:Z_k) = (0:0)\) is not even a projective point;
but even if \(e \not= 0\),
the resulting \((X_k:Z_k) = (1:0)\) may not be equal to \(\xMap([k]P)\).

In Montgomery's original context of ECM,
this is a feature, not a bug:
the ultimate goal there is to produce \((X_k,Z_k)\) 
such that \(\gcd(Z_k,N) > 1\),
regardless of whether or not
\((X_k:Z_k)\) is a correct pseudomultiplication result,
or even a legal projective point.
But it presents a complication for \(x\)-line Diffie--Hellman,
because it appears that the parties in a key exchange are obliged to
carry out some zero checks to ensure that the inputs
are not \(\xMap(O)\) or \(\xMap(T)\).

Bernstein shows that a modest modification to the map \(\xMap\) allows
such checks to be omitted entirely~\cite[Theorem 5.1]{avoidzero}.
The result is a Diffie--Hellman key exchange where the inputs are not
points on the \(x\)-line, but simple finite field elements,
by using the mapping
\(\xMap_0: \EC_{(A,B)}(\FF_q) \to \FF_q\) 
defined by
\[
    \xMap_0: P \longmapsto
    \begin{cases}
          0   & \text{if } P = O 
        \\
      x & \text{if } P = (x,y)
    \end{cases}.
\]
in place of \(\xMap\).
This means using the ladder
with input \((X_P,Z_P) = (x,1)\),
where \(x = \xMap_0(P)\) for any \(P\)
on \(\EC_{(A,B)}\) or its twist---%
so \(x\) can be any element of \(\FF_q\)---%
then returning \(x_k = X_kZ_k^{q-2}\) in \(\FF_q\)
instead of \((X_k,Z_k)\).
Note that \(x_k = X_k/Z_k\) if \(Z_k \not= 0\), and \(0\) otherwise;
in either case, \(x_k = \xMap_0([k]P)\).
Bernstein's $\xMap_0$ therefore provides pseudo-completeness for the
ladder on Montgomery curves, and an extremely simple and
efficient key exchange based on the maps \(x \mapsto x_k\).

\section{
    Differential addition chains and higher-dimensional algorithms 
}
\label{sec:higher-dim}

An \emph{addition chain} of length \(\ell\) for a nonnegative integer $k$ is an increasing
sequence of nonnegative integers, $(c_0, \dots , c_\ell)$, with $c_0=1$
and $c_\ell=k$, satisfying the following property: for all $0 < i \leq m$,
there exist $j$ and $j'$ such that $c_i = c_j+c_{j'}$ with $0 \leq j \leq
j' \leq i$. Addition chains have wide application in public key
cryptography due to their correspondence with group exponentiations. In
the context of elliptic curve cryptography, the existence of a length
$\ell$ addition chain $(c_0, \dots , c_\ell)$ for $k$ implies that the
scalar multiplication $(k,P) \mapsto [k]P$ can be computed using $\ell$
group operations. Thus, shorter addition chains require fewer
operations, and ultimately yield faster scalar multiplication routines. 

To use the fast \(x\)-line arithmetic of Montgomery curves, we need a special type of addition chain. 
A \emph{differential addition chain} of length \(\ell\) for an integer $k$ is 
a sequence $(c_0, \dots , c_{\ell+1})$ with $c_0=0$, $c_1=1$,
and (for our purposes) $k \in \{c_\ell,c_{\ell+1}\}$, together with the
property that for all $1 < i \leq \ell+1$, there exist $i'$, $j$ and
$j'$ such that $c_i = c_j+c_{j'}$ and $c_{i'} = c_{j'}-c_j$ with $0 \leq i'
\leq j \leq j' < i$. 
The existence of a length-\(\ell\) differential addition chain for \(k\)
implies that the pseudomultiplication 
\(
    (k,\xMap(P)) \mapsto \xMap([k]P)
\)
can be computed using a total of $\ell$ differential operations
(i.e., \(\xADD\)s and \(\xDBL\)s).

For an $\ell$-bit scalar, the Montgomery ladder corresponds to a
length $2\ell-1$ differential addition chain $(c_0,\dots,c_{2\ell})$; 
it requires two additions (one of which is a doubling) for each
bit of the scalar except the top bit, where only one operation is required.
In his search for shorter differential addition
chains~\cite{montgomery-lucas}, Montgomery proved that any $\ell$-bit
prime scalar requires at least 
$1.440\ell$ \(x\)-line operations, providing a lower bound on the number
of operations required in a differential addition chain in the worst
case~\cite[\S3]{montgomery-lucas}. 
(The best-case exponents, powers of 2, require 1 operation per bit.) 

\subsection{Montgomery's Euclidean algorithms}

The Montgomery ladder is a 
differential addition chain where the difference index $c_{i'}$ 
is in \(\{1,0\}\) throughout;
this corresponds to every
\(\xADD\) taking the same difference $\xMap(P)$.  
While this yields a simple and uniform algorithm, 
allowing $c_{i'}$ to vary further can yield shorter addition chains and faster scalar multiplication. 

Starting from the (subtractive) Euclidean algorithm for finding the
greatest common divisor of two integers, Montgomery derived several
algorithms for producing differential addition chains that were
significantly shorter than his ladder. The idea is to have a coprime
auxiliary exponent alongside the input~$k$, and use the
intermediate steps in the Euclidean algorithm to write down a
differential addition chain for $k$. 
This process computes a differential addition chain 
for the auxiliary exponent as well, 
so these algorithms are inherently 2-dimensional.
Here we discuss one of these algorithms, PRAC,
beginning with its 2-dimensional core
before returning to the 1-dimensional wrapper.

Algorithm~\ref{alg:EUCLID} (\EUCLID)
is a version of Montgomery's 2-dimensional PRAC subroutine
using only the ``binary'' transformations proposed by Montgomery in~\cite[Table 4]{montgomery-lucas}. 
Given a multiscalar $(m,n)$ and \(\xMap(P)\),
\(\xMap(Q)\), and \(\xMap(P\ominus Q)\), it computes \(\xMap([m]P\oplus
[n]Q)\) using only \(\xADD\) and \(\xDBL\) operations. 
This variant chain was used (for non-uniform scalar multiplications) by the implementation
of endomorphism-accelerated scalar multiplications on the 
curve in Example~\ref{eg:CHScurve}.

Since the difference arguments to the \(\xADD\)s in \EUCLID{} vary,
we must be careful about handling differences like \(\xMap(O)\) and \(\xMap(T)\),
which cause the \(\xADD\) and \(\xDBL\) we defined in
Algorithms~\ref{alg:xADD} and~\ref{alg:xDBL} to degenerate.
For simplicity, in this section we suppose that
\(\xADD\) and \(\xDBL\) have been extended to cover all inputs
(perhaps using conditional code, which is acceptable in these
non-uniform algorithms).

\begin{algorithm}
    \caption{\EUCLID: 2-dimensional scalar pseudomultiplication}
    \label{alg:EUCLID}
    \SetKwInOut{Cost}{Cost}
    \KwIn{$m,n \in \ZZ_{>0}$,
        and $\xMap(P)$, $\xMap(Q)$, and $\xMap(Q \ominus P)$
        for some $P$ and $Q$ in \(\EC_{(A,B)}(\FF_q)\)
    }
    \KwOut{$\xMap([m]P\oplus [n]Q)$}
    \SetKwData{x}{x}
    \SetKwData{mult}{s}
    \(
        \big((\mult_0,\mult_1),(\x_0,\x_1,\x_\ominus) \big)
        \gets 
        \big((m,n),(\xMap(P),\xMap(Q),\xMap(Q\ominus P))\big)
    \)
    \;
    \label{alg:EUCLID:start-while-1}
    \While{$\mult_0 \neq 0$}{
        \If{$\mult_1 < \mult_0$}{
            \(
                \big((\mult_0,\mult_1),(\x_0,\x_1,\x_\ominus)\big)
                \gets 
                \big((\mult_1,\mult_0),(\x_1,\x_0,\x_\ominus)\big)
            \)
    	}
        \uIf(\tcp*[f]{"Fibonacci" step}){$\mult_1 \leq 4\mult_0$}{
            \label{alg:EUCLID:fib-start}
            \(
                \big((\mult_0,\mult_1),(\x_0,\x_1,\x_\ominus)\big)
                \gets
                \big(
                    (\mult_0,\mult_1-\mult_0), 
                    (\xADD(\x_1,\x_0,\x_\ominus), \x_1, \x_0)
                \big)
            \)
        }
        \label{alg:EUCLID:fib-stop}
        \uElseIf{$\mult_0 \equiv \mult_1 \pmod{2}$} {
            \(
                \big((\mult_0,\mult_1),(\x_0,\x_1,\x_\ominus)\big)
                \gets
                \big((\mult_0,\frac{\mult_1-\mult_0}{2}),
                 (\xADD(\x_1,\x_0,\x_\ominus), \xDBL(\x_1), \x_\ominus)
                \big)
            \)
        }
        \uElseIf{$\mult_1 \equiv 0 \pmod{2}$} {
            \(
                \big((\mult_0,\mult_1),(\x_0,\x_1,\x_\ominus)\big)
                \gets
                \big((\mult_0,\mult_1/2),
                 (\x_0, \xDBL(\x_1), \xADD(\x_1,\x_\ominus,\x_0))
                \big)
            \)
        }
        \Else {        
            \(
                \big((\mult_0,\mult_1),(\x_0,\x_1,\x_\ominus)\big)
                \gets
                \big((\mult_0/2,\mult_1),
                 (\xDBL(\x_0), \x_1, \xADD(\x_0,\x_\ominus,\x_1))
                \big)
	        \)
	    }
    }
    \label{alg:EUCLID:end-while-1}
    \While{$\mult_1 \equiv 0 \pmod{2}$} {
        \label{alg:EUCLID:start-while-2}
        \((\mult_1,\x_1) \gets (\mult_1/2,\xDBL(\x_1)) \)
    }
    \label{alg:EUCLID:end-while-2}
    \If{$\mult_1 > 1$} {
       \( \x_1 \gets \LADDER(\mult_1,\x_1) \)   
       \;
       \label{alg:EUCLID:ladder}
    }   
    \Return{ \(\x_1\) } \tcp*{\(\x_1 = \xMap([m]P\oplus [n]Q) \) }
\end{algorithm}

At first glance, \EUCLID{} is much more complicated than \LADDER{},
but it is in fact remarkably simple and elegant.
Suppose we want to compute \([m]P\oplus[n]Q\).
Lines~\ref{alg:EUCLID:start-while-1}
through~12 
maintain the following invariants:
\SetKwData{s}{s}
\(\gcd(\s_0,\s_1) = \gcd(m,n)\)
(the reader may recognise a subtractive Euclidean algorithm here),
and
\( 
    (\x_0,\x_1,\x_\ominus) 
    = 
    (\xMap(R_0),\xMap(R_1),\xMap(R_1\ominus R_0))
\)
for some \(R_0\) and \(R_1\) in \(\EC_{(A,B)}\)
such that \([\s_0]R_0\oplus[\s_1]R_1 = [m]P\oplus[n]Q\).
Hence after the first \texttt{while} loop, 
having arrived at \(\s_0 = 0\),
we must have \(\s_1 = \gcd(m,n)\)
and \(\x_1 = \xMap([m/\s_1]P\oplus[n/\s_1]Q)\).
To complete the task,
it suffices to carry out a 1-dimensional pseudomultiplication
of \(\x_1\) by \(\s_1 = \gcd(m,n)\)
(in Line~\ref{alg:EUCLID:ladder}
we use~\LADDER); 
of course, if \(m\) and~\(n\) are random,
then we expect \(\s_1\) to be quite small at this point.
The second \texttt{while} loop
(Lines~\ref{alg:EUCLID:start-while-2}-14)
slightly optimizes this final 1-dimensional pseudomultiplication
by using pure pseudo-doubling to exhaust any power of~2 in~\(\s_1\),
saving a few superfluous \xADD{}s in the \LADDER{} call.

If we had an efficient pseudo-tripling operation $\xMap(P)
\mapsto \xMap([3]P)$ are relatively efficient, it would be advantageous
to include Montgomery's ``ternary'' transformations in the loop of Algorithm~\ref{alg:EUCLID}. 
Independent analyses by Stam~\cite[Conjecture 3.29]{stamthesis} and
Bernstein~\cite[\S3]{DJB-chain} conclude that
this full version of Montgomery's 2-dimensional PRAC subroutine 
computes $\xMap([m]P \oplus [n]Q)$ 
for $\ell$-bit multiscalars $(m,n)$
using an average total of $1.82\ell$ differential operations.

Returning to the 1-dimensional problem,
Montgomery suggests computing
\(\xMap([k]P)\)
as 
\(
 \xMap([m]P\oplus [n]Q) 
\), 
where \((m,n)\) and $Q = [\lambda]P$ 
satisfy $[m+n \lambda]P = [k]P$ 
for some \(\lambda\) chosen to minimize the complexity of the
pseudomultiplication.
If we can precompute \(\xMap([\lambda]P)\) from $\xMap(P)$,
then any $\ell$-bit pseudomultiplication
can be performed as an $\ell/2$-bit double-pseudomultiplication,
and the resulting differential addition chains 
have average length $0.92\ell$~\cite[Corollary 3.32]{stamthesis}. 
But without precomputation,
and following his heuristics for choosing $\lambda$,
Montgomery's PRAC algorithm produces differential addition
chains of average length $1.64\ell$
for \(\ell\)-bit scalars~\cite[Corollary 3.35]{stamthesis}.

Algorithm~\ref{alg:PRAC}
is a simplified version of PRAC:
we omit the repeated-tripling step that,
like the ternary steps omitted in its subroutine
Algorithm~\ref{alg:EUCLID},
presupposes the existence of rapid pseudo-tripling on \(\EC_{(A,B)}\).
The idea of taking 
\(
    (m,n) 
    =
    (\lfloor{k/\varphi}\rfloor,k-\lfloor{k/\varphi}\rfloor)
\),
where \(\varphi = (1 + \sqrt{5})/2\) is the famous golden ratio,
is that this induces a sequence 
of roughly \((\frac{1}{2}\log_\varphi{2})\ell\)
of the relatively cheap so-called Fibonacci branch 
(Lines~\ref{alg:EUCLID:fib-start}-6) 
of Algorithm~\ref{alg:EUCLID},
followed by roughly \(\frac{1}{2}\ell\)
branches distributed as for random multiscalars.
We refer the reader to Stam's thesis~\cite[\S3.3.4]{stamthesis}
for further details and analysis.

\begin{algorithm}
    \caption{\PRAC: (simplified) 1-D Euclidean pseudomultiplication}
    \label{alg:PRAC}
    \SetKwInOut{Cost}{Cost}
    \KwIn{$k \in \ZZ_{>0}$, 
        and $\xMap(P)$ 
        for some $P$ in \(\EC_{(A,B)}(\FF_q)\) 
    } 
    \KwOut{$\xMap([k]P)$}
    \SetKwData{x}{x}
    \SetKwData{r}{r}
    \SetKwData{mult}{s}
    \((\mult,\x) \gets (k,\xMap(P))\)
    \;
    \While{\(\mult \equiv 0 \pmod{2}\)}{ 
        \((\mult,\x) \gets (\mult/2,\xDBL(\x))\)
    }
    \(\r \gets \lfloor{\mult/\varphi}\rfloor\) \textbf{where} \(\varphi = (1 + \sqrt{5})/2\)
    \;
    \(\x \gets \EUCLID((\r,\mult-\r),(\x,\x,\xMap(0)))\)
    \;
    \Return{\x}
\end{algorithm}

\subsection{Higher-dimensional ladder analogues}

Although the short addition chains produced by Montgomery's Euclidean
algorithms are suitable for his original application to ECM, their
non-uniform and variable-time behavior makes them less suitable for
application to ECC, where uniformity is a mandatory first step towards
hiding secret exponents from adversaries exploiting side-channels. 
For cryptographic multiscalar multiplications, 
we may need higher-dimensional differential addition chains that share the uniform behavior of the 1-dimensional Montgomery ladder. 

Bernstein defines a 2-dimensional analogue of the Montgomery
ladder in~\cite{DJB-chain}
(the ``binary chain''). 
Given an \(\ell\)-bit multiscalar \((m,n)\) 
and the values of \(\xMap(P)\), \(\xMap(Q)\), \(\xMap(P\ominus Q)\) and \(\xMap(P\oplus Q)\), this algorithm computes
\(\xMap([m]P\oplus[n]Q)\) using exactly \(3\ell\) differential operations.
If \(\xDBL\) and \(\xADD\) are implemented in a uniform way,
then Bernstein's algorithm is also uniform: 
it is a sequence of \(\ell\) 
steps, each consisting of one \(\xDBL\) and two \(\xADD\)s.
Further, in each of those steps,
the argument of \(\xDBL\) is shared by one of the \(\xADD\)s,
and in some contexts these calls may be merged to use some shared
computations. 

Every call to \(\xADD\) in Bernstein's algorithm takes 
one of the four input values \(\xMap(P)\), \(\xMap(Q)\),
\(\xMap(P\ominus Q)\) and \(\xMap(P\oplus Q)\) 
as its difference argument; 
this makes it possible
to recover the correct $y$-coordinate for the output \(\xMap([m]P\oplus[n]Q)\)
using exactly the same technique as in~\S\ref{sec:recovery}. 
This makes Bernstein's chain a viable option
for computing the multiscalar multiplication during the verification
phase of signature schemes that take advantage of differential
arithmetic.

Brown defined an \(n\)-dimensional analogue of the Montgomery ladder
in~\cite{Brown}. Given an $n$-dimensional multiscalar \((m_1,\dots,m_n)\)
in \(\ZZ^n\), Brown's algorithm uses \(x\)-line arithmetic to
compute $\xMap([m_1]P_1\oplus\dots\oplus[m_n]P_n)$ from the $(3^n-1)/2$ input
values $\xMap([e_1]P_1\oplus\dots\oplus[e_n]P_n)$ with the $e_i \in \{-1,0,1\}$ and not all zero.

\begin{example}[A Montgomery $\mathbb{Q}$-curve reduction~\cite{CHS}]
    \label{eg:CHScurve}
    The Montgomery curve
    \[
        \EC: y^2 = x(x^2 + Ax + 1)
        \quad
        \text{over}
        \quad
        \ \FF_{p^2} = \FF_{2^{127}-1}(\sqrt{-1})
    \]       
    with
    \(
        A  = 45116554344555875085017627593321485421 + 2415910908\sqrt{-1}
    \)
    has \(\#\EC(\FF_p) = 4r\) 
    and \(\#\EC'(\FF_p) = 8r'\), where $r$ and $r'$
    are 252- and 251-bit primes, respectively. 
    In~\cite{CHS} we see that 
    \(\EC\) is equipped with an efficiently computable endomorphism \(\psi\)
    of degree~\(2p\),
    which acts on \(\EC(\FF_{p^2})[r]\)
    as \([\lambda]\) where \(\lambda^2 \equiv -2\pmod{r}\).
    The classic Gallant--Lambert--Vanstone (GLV) technique~\cite{GLV01}
    could be used to compute 1-dimensional scalar multipications 
    \([k]P\) as 2-dimensional multiplications \([m]P\oplus[n](\psi(P))\),
    with \(m\) and \(n\) of roughly 128 bits.
    The scalar multiplication implementation in~\cite{CHS}
    transports the GLV approach to the \(x\)-only setting,
    simultaneously exploiting Montgomery arithmetic.
    First, we compute \(\xMap(P) \mapsto \xMap(\psi(P))\)
    and \(\xMap(P) \mapsto \xMap((\psi-1)P) = \xMap(\psi(P)\ominus P)\).
    Then given any \(k\) in \([0,r)\)
    we can compute \(\xMap([k]P) =
    \xMap([m]P\oplus[n]\psi(P))\),
    where \(m\) and \(n\) are 128-bit scalars,
    using Algorithm~\ref{alg:EUCLID} for public scalars
    and Bernstein's uniform 2-dimensional binary chain for private scalars.
\end{example}

\section{
    Generalizations and other applications
}
\label{sec:HECC}

The Montgomery ladder is a general algorithm that works in any abelian
group, and in group quotients where analogues of \(\xADD\) are available.
The most interesting groups of this kind---at least from a cryptographic
point of view---are other models of elliptic
curves, and Jacobians of hyperelliptic curves.

\subsection{The Montgomery ladder on other models of elliptic curves}

The Montgomery ladder can be applied to any model of an elliptic curve,
using either the usual addition and doubling operations, or the analogue
of \(x\)-only arithmetic.
For Weierstrass models,
the chief interest in the Montgomery ladder is its side-channel
resistance when computing scalar multiples with secret scalars,
as explored by Brier and Joye
in~\cite{Brier--Joye}. Gaudry and Lubicz~\cite{gaudry-lubicz} used the
classical complex-analytic theory of theta functions to derive
pseudo-group law formul\ae{} for $\PP^1 \cong \EC_\lambda /
\langle \ominus \rangle$, where $\EC_\lambda \colon
y^2=x(x-1)(x-\lambda)$ is a Legendre model. Their
formul\ae{} offer trade-offs with the $\xADD$ and $\xDBL$ operations on Montgomery curves, which could make them favorable in certain scenarios.

Castryck, Galbraith, and Farashahi~\cite{CGF08} made the striking
suggestion of using Montgomery curves \emph{in conjunction} with
their corresponding twisted Edwards models.
The transformations of~\eqref{eq:Montgomery-to-Edwards}
and~\eqref{eq:Edwards-to-Montgomery}
are extremely easy to compute;
this allows a mixed arithmetic, 
passing back-and-forth between 
$x$-only pseudo-additions on the Montgomery curve 
and $v$-only pseudo-doublings on the corresponding Edwards curve. 

\subsection{The Montgomery ladder on hyperelliptic curves}

The Montgomery ladder has been applied with great success in hyperelliptic
cryptography based on genus-2 curves.
The story begins with Smart and Siksek~\cite{smart-siksek},
who observed that we can use pseudo-additions to instantiate Diffie--Hellman key exchange on Kummer surfaces
(quotients of Jacobians of genus-2 curves by \(\pm 1\)). 
Duquesne~\cite{Duquesne04} made this concrete by
combining the ladder with explicit
formul\ae{} for arithmetic on genus-2 curves
of the form \(\HC: y^2 = xf(x)\), 
where \(f\) is a squarefree degree-4 polynomial.
While the factor of \(x\) on the right-hand-side
also appears in the defining equations of Montgomery curves, 
Duquesne's curves are not a true genus-2 Montgomery analogue:
we would expect a particularly simple action on the Jacobian
of the elements of the kernel of a \((2,2)\)-isogeny factoring
\([2]\), for example, mirroring the special form of the
translation-by-\(T\) map on Montgomery curves, but no such structure
is imposed by Duquesne's form.  Nevertheless, this special curve form
allows a small speedup over general genus-2 arithmetic.
Duquesne later described the Montgomery
ladder on Kummer surfaces of arbitrary genus-2 curves~\cite{Duquesne10},
building on Flynn's arithmetic of general Kummer
surfaces~\cite[Chapter 3]{casselsflynn}.

Gaudry used the Montgomery ladder for efficient pseudomultiplication on
a model of the Kummer
with especially fast pseudo-addition and pseudo-doubling operations~\cite{Gaudry}, 
building on observations of
D.~V. and G.~V.~Chudnovsky~\cite{chudnovsky86}.
Here, the efficiency really is a consequence 
of a special \(2\)-torsion structure:
all of the \(2\)-torsion points are defined over \(\FF_q\),
and their translations act linearly on the Kummer 
by diagonal and permutation matrices.
This approach has successfully used in high-speed Diffie--Hellman
implementations~\cite{BoCoHi13,kummerstrikes,muKummer},
and a hyperelliptic generalization of 
ECM factorization~\cite{Cosset10}.

An analogue of \(y\)-coordinate recovery 
(see \S\ref{sec:recovery}) exists in genus 2:
Chung and the authors give an explicit algorithm in~\cite{CCS},
recovering Jacobian elements from the output of the Montgomery ladder
on the Kummer.
This enables the implementation of full signature schemes
using Kummer surfaces~\cite{muKummer}.

All of these ideas and techniques are carried much further in the setting of
higher-dimensional abelian and Kummer varieties
by Lubicz and Robert~\cite{Lubicz--Robert}. 

\bibliography{bib} 
\bibliographystyle{plain}

\end{document}